\documentclass[reprint,twocolumn,secnumarabic,amssymb, nobibnotes, aps, pra,superscriptaddress]{revtex4-1}

\usepackage{extarrows}
\usepackage{amsmath}    
\usepackage{graphicx}   
\usepackage{verbatim}   
\usepackage{color}      
\usepackage{subfigure}  
\usepackage{hyperref}   
\usepackage{verbatim}   
\usepackage[normalem]{ulem}
\begin{document}


\title{X-ray spectroscopy of rare-earth nickelate LuNiO$_3$:~LDA+DMFT study}

\author{Mathias Winder}
\affiliation{Institute of Solid State Physics, TU Wien, 1040 Vienna, Austria}
\author{Atsushi Hariki}
\thanks{hariki@osakafu-u.ac.jp}
\affiliation{Institute of Solid State Physics, TU Wien, 1040 Vienna, Austria}
\affiliation{Department of Physics and Electronics, Graduate School of Engineering, Osaka Prefecture University 1-1 Gakuen-cho, Nakaku, Sakai, Osaka 599-8531, Japan}
\author{Jan Kune\v{s}}
\thanks{kunes@ifp.tuwien.ac.at}
\affiliation{Institute of Solid State Physics, TU Wien, 1040 Vienna, Austria}
\affiliation{Institute of Physics, Czech Academy of Sciences, Na Slovance 2, 182 21 Praha 8, Czechia}

\date{\today}
\begin{abstract}
We present a computational study of resonant inelastic x-ray scattering (RIXS) and
x-ray absorption in a representative rare-earth nickelate LuNiO$_3$. We study the
changes in the spectra across the metal-insulator/site-disproportionation transition.
In particular, we address the question of site-selectivity of the two methods in
the disproportionated insulating phase and the signature of metal-insulator transition
in the fluorescence-like feature of the RIXS spectra. To this end we use 
the local density approximation + dynamical mean-field theory (LDA+DMFT) approach
combined with configuration integration method to compute the core-level spectra.

\end{abstract}

\maketitle

Core-level spectroscopies present a powerful set of tools for investigation of solids~\cite{groot_kotani}.
The strong energy dependence of the x-ray absorption edges on proton number makes the contributions of
different elements easy to distinguish.
In fact, typical spectral shapes and eV-scale shifts allow the x-ray absorption spectroscopy (XAS) to distinguish and identify different valence states of the same element.
Site-selectivity is particularly useful for investigation of disproportionation phenomena, in which inequivalent atoms of the same element appear spontaneously.
A much studied example are rare-earth nickelates RNiO$_3$, which exhibit a thermally driven simultaneous
structural and metal-insulator transition accompanied by a charge disproportionation (CD) on Ni sites~\cite{Lu18,Fursich19,Green16,Mazin07,Lee11,Park12,Lau13,Johnston14,Jaramillo14,Subedi15,Bisogni16,Mercy17,Varignon17,Ruppen15}.

Formation of an immobile excitonic state 
between the core-hole and excited $d$-electron limits the information about the ground state and its low-energy excitations that can be extracted from XAS spectra~\cite{groot_kotani,Groot90}.
Resonant inelastic x-ray scattering (RIXS) resolves this deficiency and  opens a new route to study (charge-neutral) two-particle (2P) excitations in
CD materials~\cite{Lu18,Elnaggar18,Elnaggar19,Fursich19}.
The price is the necessity of theoretical simulations to interpret the spectra arising from the complex RIXS excitation process~\cite{Kotani01,ament11,groot_kotani}.
Similar to XAS, RIXS is element-selective by setting the incoming photon energy $\omega_{\rm in}$ at the element-specific absorption edge.
Site-selective interpretation of RIXS spectra, proposed in some CD materials~\cite{Lu18,Elnaggar18,Elnaggar19,Fursich19},
remains an open question.



A double peak at the Ni $L_3$-edge of XAS spectra of 
RNiO$_3$ was recently associated with two distinct Ni sites in the CD phase by a theoretical analysis using a double-cluster model~\cite{Green16}.
Bisogni $et$ $al$.~\cite{Bisogni16} reported a high-resolution RIXS across the $L_3$ double-peak, which revealed an unusual coexistence of Raman-like (RL) and fluorescence-like (FL) features.
The behavior of the FL feature across the transition between the CD-insulating and metallic phases was interpreted as a signature of metal-insulator transition~\cite{Bisogni16}.
This result, 
as well as site-selectivity achieved by tuning the incoming photon energy
$\omega_{\rm in}$~\cite{Bisogni16,Lu18,Fursich19},
calls for deeper theoretical investigation.

In this Letter, we study Ni $L_3$ RIXS and XAS spectra of a representative nickelate LuNiO$_3$ across
the transition from paramagnetic metal (PMM) to 
paramagnetic insulator (PMI) at $T_{\rm MI} \sim 600$~K
characteristic for the RNiO$_3$ family~\cite{Catalano18,Torrance92,Alonso99,Catalan08}.
To this end we use local-density approximation (LDA) + dynamical mean-field theory (DMFT)~\cite{georges96,kotliar06,kunes09} augmented with Anderson impurity model (AIM) description of the core-level spectra~\cite{Hariki17,Hariki18,Hariki20,Ghiasi19,Hariki20,Jindrich18}.
The method provides a computationally feasible description of RIXS, which includes self-consistently coupled inequivalent Ni ions
as well as the electron-hole continuum of the extended system.  

\begin{figure}[t] 
\includegraphics[width=1.00\columnwidth]{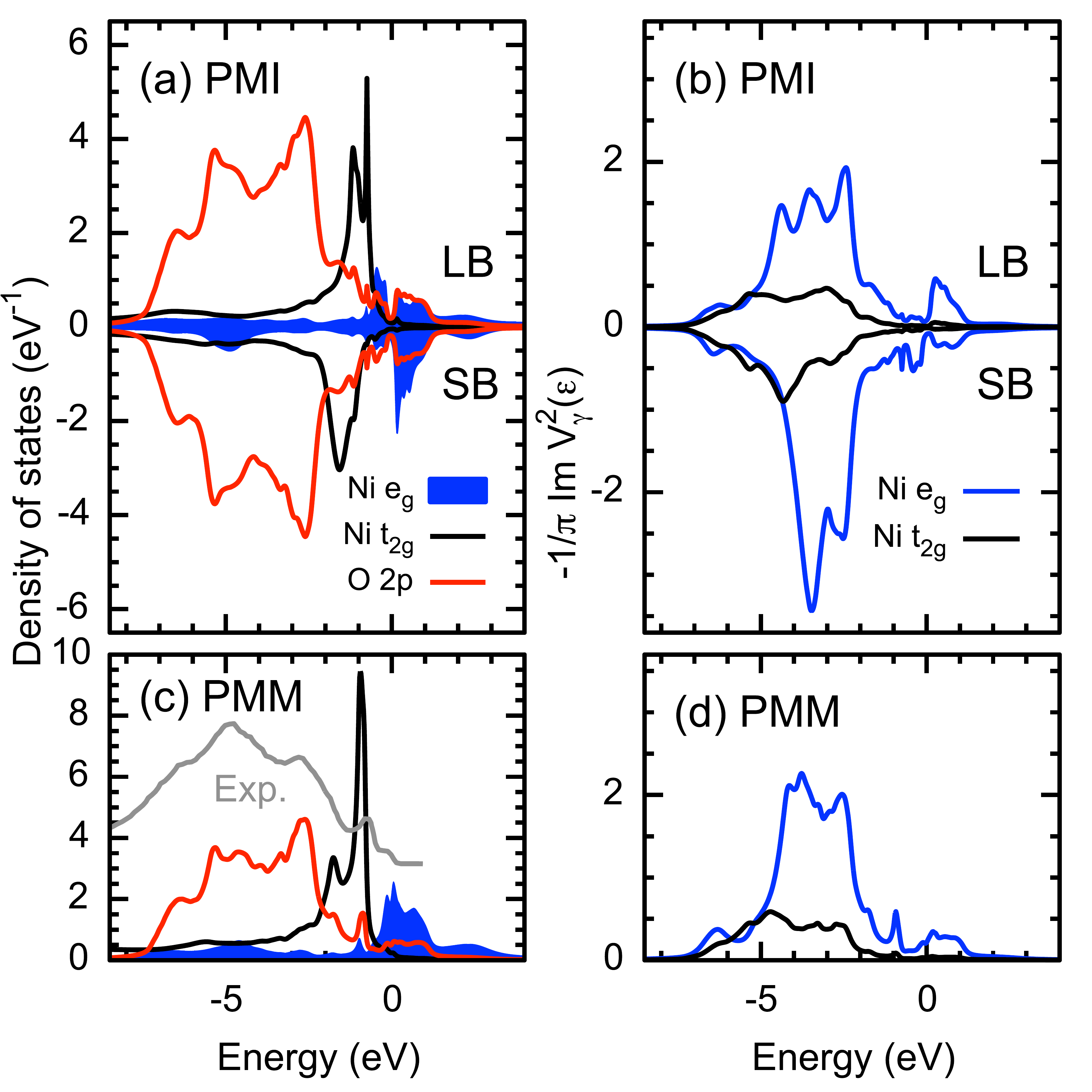}
\caption{The 1P spectra calculated by LDA+DMFT in (a) PMI phase and (c) PMM phase with experimental valence photoemission spectrum (Ref~\cite{Eguchi09}).
DMFT hybridization intensities $V^2_{S}(\varepsilon)$ in (b) PMI phase and (d) PMM phase.
In the PMI phase, the 1P spectra and $V^2_{S}(\varepsilon)$ at the Ni SB site are multiplied by $-1$.
The $\mu_{\rm dc}$-dependence of the 1P spectra can be found in SM~\cite{sm}.}
\label{fig_dos}
\end{figure}

The 
calculation 
proceeds in following steps~\cite{Hariki17,Hariki18,Hariki20}.
First, 
the LDA bands for high ($Pbnm$, 533~K) and low ($P2_1/n$, 673~K) temperature structures~\cite{Alonso01} are obtained using Wien2K package~\cite{wien2k}, and subsequently are projected onto a $dp$ tight-binding model spanning Ni 3$d$ and O 2$p$ orbitals~\cite{wien2wannier,wannier90}.
The bare energy of the Ni $3d$ states is obtained from the LDA values by subtracting the so called double-counting correction $\mu_{\rm{dc}}$, which accounts for the $dd$ interaction
present in the LDA calculation. In absence of a unique definition of $\mu_{\rm{dc}}$~\cite{kotliar06,karolak10}, we treat it
as a parameter~\cite{Hariki17} adjusted by comparison to the experimental valence photoemission spectra, see Fig.~\ref{fig_dos}c and SM~\cite{sm} for details.
The low-temperature structure contains two inequivalent Ni sites with long (short) Ni--O bonds, referred as LB (SB) site.  
The $dp$ model is augmented with the local interaction within the Ni 3$d$ shells, with the Coulomb $U$ and Hund's $J$ parameters 
$(U,J)$=(7.0~eV, 0.8~eV) adopted form previous LDA+DMFT studies
~\cite{Nowadnick15,Haule17}.
The DMFT, employing strong-coupling continuous-time quantum Monte Carlo (CT-QMC) impurity solver~\cite{werner06,boehnke11,hafermann12,Hariki15},
is used to obtain the site-dependent hybridization function $V_{S}(i\omega_n)$ for each Ni site~\cite{georges96,Park12,sm}.
Construction of the AIM is concluded
by analytic continuation of $V_{S}(\varepsilon)$ to real frequencies~\cite{jarrell96}.


Next, we augment the AIM for each Ni site with the $2p$ orbitals and $2p$--$3d$ interaction~\footnote{The 2$p$-3$d$ Coulomb parameters for the anisotropic part ($F^2$, $G^1$, $G^3$)~\cite{groot_kotani,slater} are calculated by an atomic Hartree-Fock calculation. The $F^2$, $G^1$ and $G^3$ values are scaled down to 75\% of their actual values to simulate the effect of intra-atomic configuration interaction from higher basis configurations neglected in the atomic calculation~\cite{cowan,Sugar72,Tanaka92,Matsubara05}.
We fix $U_{dc}=1.3\times U_{dd}$~\cite{bocquet92,park88,Hariki17}, where $U_{dc}$ ($U_{dd}$) is the configuration-averaged Coulomb parameter in the Ni 2$p$--3$d$ (3$d$--3$d$) interaction.
The explicit form of the AIM Hamiltonian is given in SM and Refs.~\cite{Hariki17,Hariki18,Ghiasi19}.}.
The RIXS spectrum is calculated as a sum of site-contribution 
obtained by the Kramers-Heisenberg formula~\cite{Kramers25,groot_kotani} for the respective AIM
\begin{align}
\label{eq:siterixs}
F^{S}_{\rm RIXS}(\omega_{\rm out},\omega_{\rm in})&=
     \sum_{f,n} \left| \langle f | T_{\rm e} 
     \frac{1}{\omega_{\rm in}+E_n-\hat{H}_{\rm AIM}^{(S)}+i\Gamma}
     T_{\rm i} | n \rangle \right|^2 \notag \\
     &\times e^{-\beta E_n}/Z \times \delta(\omega_{\rm in}+E_n-\omega_{\rm out}-E_f). 
\end{align}
Here, $|n\rangle$ and $|f\rangle$ are 
eigenstates of AIM with energies $E_n$ and $E_f$. 
Similarly, the XAS sectra are obtained by summing the site contributions, see SM~\cite{sm}.
In this approach, the interference between processes on different sites is neglected.

A configuration-interaction (CI) solver is employed to compute RIXS intensities in Eq.~(\ref{eq:siterixs})~\cite{Hariki17,Hariki18}.
We used 30~bath levels per impurity spin and orbital.
We checked that the CI solver reproduces well the reduced density matrix obtained by the CT-QMC simulation (with the continuum bath).
We refer the reader interested in the technical details to SM~\cite{sm}.

\setlength{\tabcolsep}{0.77em}
\begin{table}[t]
\begin{tabular}{ c | r r r | r r r }
\toprule
& \multicolumn{3}{c}{CI} &  \multicolumn{3}{c}{CT-QMC}\\
\colrule
Sector & LB   & SB & PMM  & LB   & SB & PMM
\\
\colrule
$d^6$ & 0.3 & 2.4    & 0.4     & 0.4 & 2.0   & 1.0  \\
$d^7$ & 17.9 & 34.3  & 23.2    & 17.6 & 31.1 & 25.5 \\
$d^8$ & 70.9 & 52.7  & 65.2    & 69.5 & 54.4 & 60.5 \\
$d^9$ & 10.7 & 10.3  & 11.0    & 12.2 & 12.1 & 12.5 \\
$d^{10}$ & 0.2 & 0.3 & 0.2     & 0.3 & 0.5   & 0.3  \\
\botrule
\end{tabular}
\caption{The atomic weights (diagonal elements of the site-reduced density matrix) integrated in the $N= 6 \sim 10$ sectors.
The weights are calculated on the AIM of the LB site and the SB site, and the PMM phase. The numbers estimated by the CI solver (descritized bath) and the CT-QMC solver (continuum bath) are compared. The CI expansion scheme used in the estimate is summarized in SM~\cite{sm}.} 
\label{tab:weights} 
\end{table}

\begin{figure}[t] 
\includegraphics[width=0.98\columnwidth]{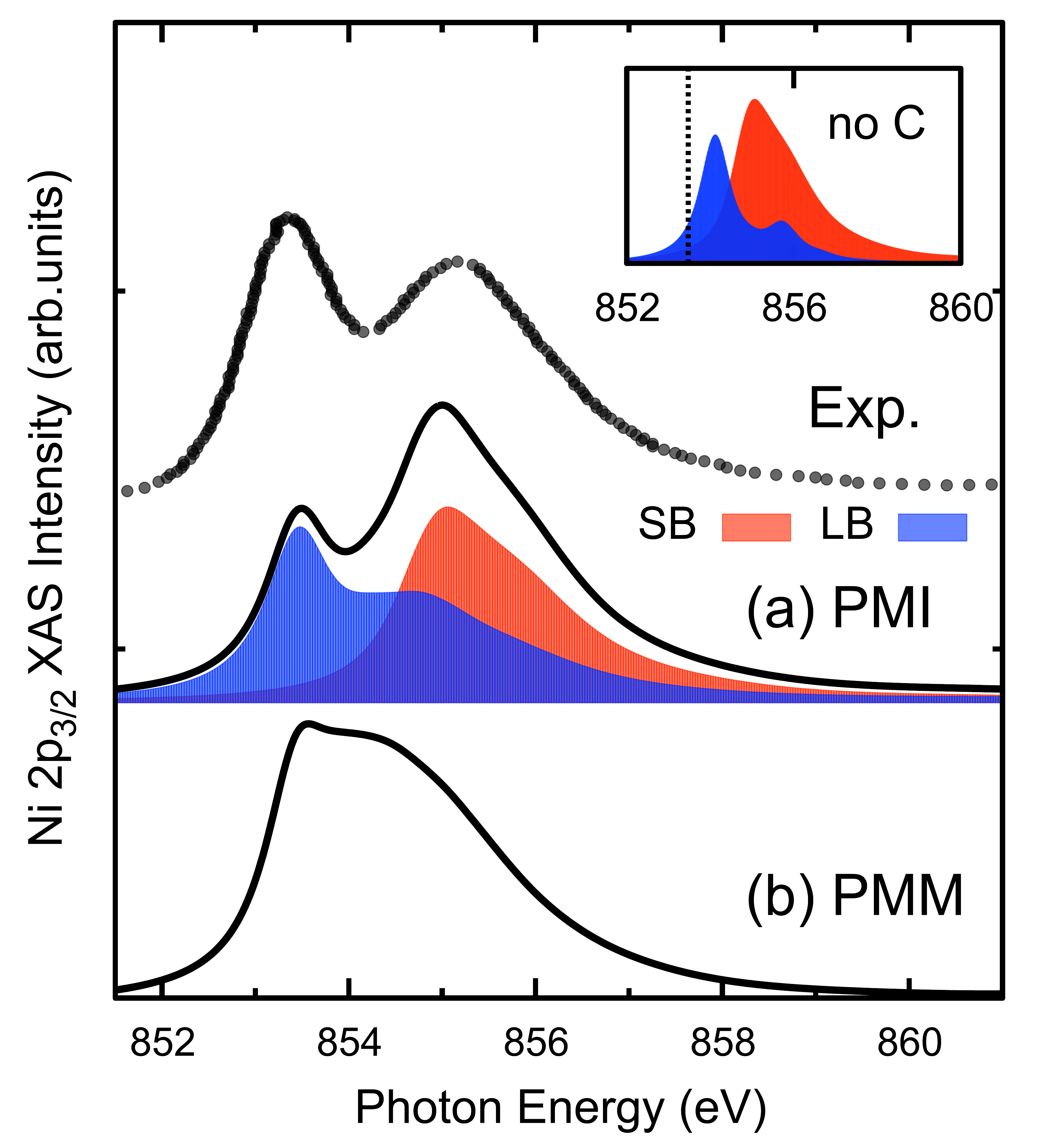}
\caption{Ni $L_3$-XAS in LuNiO$_3$ calculated by LDA+DMFT in (a) PMI phase and. (b) PMM phase.
In the PMI phase, LB (SB) Ni site contribution is shown in blue (red) color.   
The experimental Ni $L_3$-XAS data (Ref.~\cite{Piamonteze05}) measured in the PMI phase is shown together.
The spectral broadening is considered using a Lorentzian 400~meV (HWHM).
The $\mu_{\rm dc}$-dependence of the XAS spectra can be found in SM~\cite{sm}. 
The inset shows the site contribution calculated with no a hybridization above $E_F$ in the XAS final states, see details in texts. The horizontal dashed line marks the peak position of the LB XAS spectrum in the panel~(a).}
\label{fig_xas}
\end{figure}

\begin{figure}[t] 
\includegraphics[width=0.95\columnwidth]{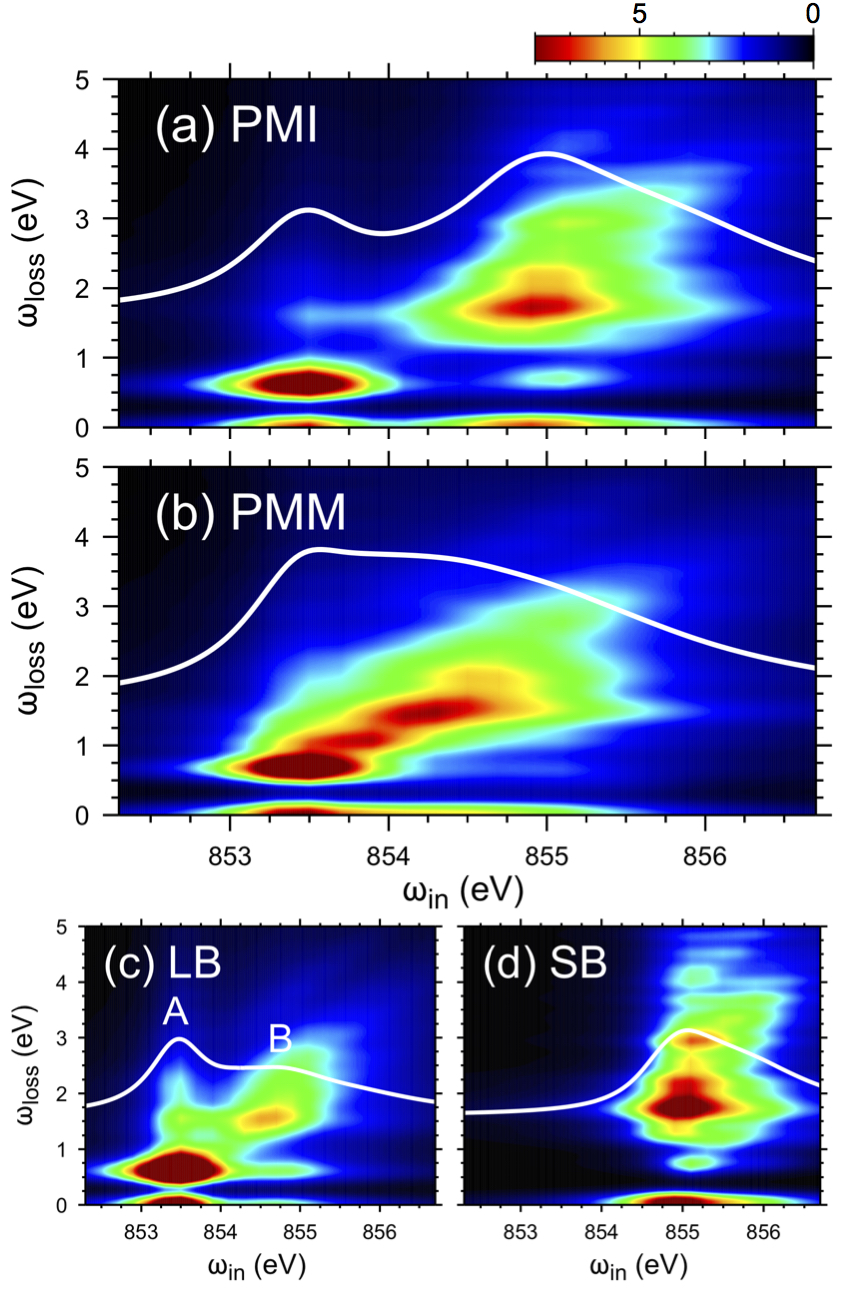}
\caption{Ni $L_3$-RIXS in LuNiO$_3$ calculated by LDA+DMFT in (a) PMI phase and (b) PMM phase.
(c) LB and (d) SB Ni site contribution to the RIXS spectra in the PMI phase.
White lines in panels are Ni $L_3$-XAS spectra calculated in the corresponding phase or Ni site.
The $\mu_{\rm dc}$-dependence of the RIXS spectra can be found in SM~\cite{sm}.
The spectral broadening is considered using a Gaussian 150~meV (HWHM).}
\label{fig_rixs}
\end{figure}

\begin{figure}[t] 
\includegraphics[width=1.0\columnwidth]{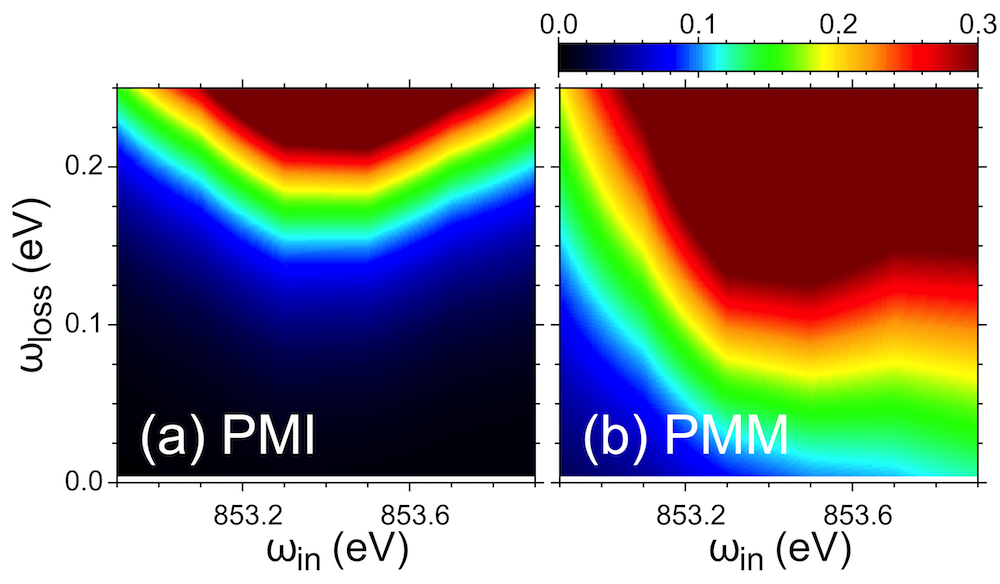}
\caption{Low-energy Ni $L_3$-RIXS features in LuNiO$_3$ in (a) PMI phase and (b) PMM phase. Note that the calculated elastic peak ($\omega_{\rm loss}$=0.0~eV) is removed.}
\label{fig_rixs_lowe}
\end{figure}


Figures~\ref{fig_dos}ac show the LDA+DMFT one-particle (1P) spectral 
densities in the PMI and PMM phases.
The gross features of the spectra in the two phases are similar spectra to one another.
The Ni $3d$ peak at $-1.0$~eV and O $2p$ peaks at $-2.5$~eV and $-5.0$~eV, corresponding to non- and anti-bonding states, respectively, match well with the experimental spectra of related nickelate LaNiO$_3$~\cite{Eguchi09}, see Fig.~\ref{fig_dos}c.
%
%
%
%
%
The hybridization densities $V_{\gamma,S}(\varepsilon)$, the amplitude
of electron hopping between the Ni $3d$ orbital $\gamma$ and
the rest of the system at energy $\varepsilon$~\footnote{Since the hybridization intensity is defined via the inverse of the local Green's function, the valence spectrum and $V_{\gamma}^2(\varepsilon)$ are not proportional each other in general.},
are shown in Figs.~\ref{fig_dos}bd.
The top of the valence band (bottom of the conduction band) is dominated by Ni LB (SB) 3$d$ $e_g$ states in Fig.~\ref{fig_dos}a, although the total Ni $d$-occupation is almost identical on the two sites.
The stronger Ni--O bonds at the SB site pushes the anti-bonding states above $E_F$, leading the sizable CD in at low energies, that is compensated by its bonding counterpart at higher energies (around $-8$~eV to $-4$~eV).
Somewhat counter-intuitively, we find larger
$V_S$ above $E_F$ for LB than for SB site, which follows from 
the overall low-energy behavior of $V_{e_g,\rm LB}$ reflecting the 1P spectral density on the SB site and vice versa.
This behavior of the hybridization function is essential for the understanding of XAS and RIXS spectra.

Application of the CI solver to metallic systems raises questions concerning discretization of the bath in AIM, reference state of the impurity or the size of the CI basis (degree of expansion). We discuss these issues in SM.
While the choice of the impurity reference state is crucial to minimize the computational effort (degree of CI expansion),
we have checked that the same impurity dynamics (spectra) is obtained for different choices.
In the present study we combine the spectra obtained from several impurity reference states (corresponding to different charge sectors of the discretized AIM).
To benchmark the CI results, we compare the equilibrium local density matrices (abundances 
of different Ni valence states) obtained with CI to those from CT-QMC in 
Table~\ref{tab:weights}.
The Ni LB site is dominated by $d^8$ ($S=1$) state, expected in ionic Ni$^{2+}$.
The SB site exhibits pronounced charge fluctuations between
$d^8$ ($\sim 54~\%$) and $d^7$ ($\sim 31~\%$) states.
The ground state of $H^{\rm{SB}}_{\rm{AIM}}$ is a spin singlet.
This behavior matches well with the notion of site-selective Mott state in RNiO$_3$~\cite{Park12,Subedi15,Ruppen15}.

Fig.~\ref{fig_xas} shows the calculated Ni $L_3$-XAS spectra in the PMI and PMM phases, together with the experimental data in the PMI phase of LuNiO$_3$~\cite{Piamonteze05}.
The experimental data exhibit a double-peak shape, composed of a sharp peak at low energy $\omega_{\rm in}$ ($\sim 853.5$~eV) and a broader feature in high $\omega_{\rm in}$ ($\sim 855$~eV). The LDA+DMFT spectra match the experiment quite well. Note that no 
by-hand alignment of LB and SB spectra was applied.
The low-$\omega_{\rm in}$ peak originates from a $2p-3d$ exciton on the LB side.
The broader high-$\omega_{\rm in}$ peak originates from
the excited $d$ electron delocalized to the bath orbitals.
This process is dominant on the SB site, but has a sizable contribution on the LB site as well. The delocalization of the excited electron from 
the LB site is facilitated by the large hybridization intensity above $E_F$ (Fig.~\ref{fig_dos}b).
This is demonstrated in the inset of Fig.~\ref{fig_xas} by setting $V_{\rm S}(\varepsilon>0)$ to zero.
There, the broad high-$\omega_{\rm in}$ feature in the LB spectra vanishes.
The small peak, that remains at $856$~eV is due to a core-valence multiplet as observed in typical Ni$^{2+}$ insulators like NiO~\cite{Alders98,Groot90,groot_kotani}.
The overlap of the LB and SB signals limits the site-selectivity 
of $2p$ XAS in RNiO$_3$ nickelates.



Figs.~\ref{fig_rixs}ab show the calculated Ni $L_3$-RIXS spectra in the PMI and PMM phases, respectively. 
The results reproduce well the experimental observations by Bisogni $et$ $al$.~\cite{Bisogni16}:~the RL feature ($\omega_{\rm loss}\sim$~1~eV) with a constant emission energy $\omega_{\rm loss}$ coexists with the FL feature showing a constant $\omega_{\rm out}$ behavior, i.e.,~a linear dependence of $\omega_{\rm loss}$ on $\omega_{\rm in}$. 
In the PMI phase, the FL intensity is slightly suppressed at $\omega_{\rm in}$ between the $L_3$ double-peak (i.e.,~at around $\omega_{\rm in}=854$~eV) compared to the one in the PMM phase, which is also observed in the experiment~\cite{Bisogni16}. 
Figs.~\ref{fig_rixs}cd show the site-resolved RIXS spectra in the PMI phase.
As we mentioned above, the LB $L_3$-XAS consists of two features marked as $A$ (excitonic peak) and $B$ (continuum). Their different character is reflected in the 
$\omega_{\rm in}$-dependence of the LB RIXS spectra:
~RL feature (due to inter-atomic $dd$ excitations) resonates mainly at $A$, while the FL feature due to unbound electron-hole pairs gains intensity with approaching $B$.
The latter can be understood as the x-ray excited electron, which leaves the LB site in the intermediate state of RIXS, giving rise to an unbound electron-hole pair in the RIXS final state (Fig.~\ref{fig_rixs}c). The propensity of the excited electron to escape the TM site is encoded in the
hybridization intensity above the Fermi level.
The SB signals in Fig.~\ref{fig_rixs}d shows an intense CT excitation extending to  higher $\omega_{\rm loss}$ in addition to a less prominent FL feature, which merges with the CT excitations at $\omega_{\rm loss}\sim 3$~eV.
The bright and (vertically) broad CT feature 
reflects stronger Ni--O hybridization on the SB site. 
Since the $B$-peak signal from the LB site
largely overlaps with the SB signal, Fig.~\ref{fig_xas}, the site-selectivity cannot be achieved for the corresponding $\omega_{\rm{in}}$.
In RNiO$_3$, only the $d$--$d$ excitations at the A-peak can be associated with the LB site.



Finally, we zoom at very low $\omega_{\rm loss}$ to address the behavior reported in Ref.~\onlinecite{Bisogni16}.
Figs.~\ref{fig_rixs_lowe}ab show the calculated low-$\omega_{\rm loss}$ RIXS spectra in the PMI phase and the PMM phase, respectively.
The photon energies $\omega_{\rm in}$ are set to around the main peak at the LB site.
We find a significant difference, gap closing, between the CD-insulating and metallic phases, which matches the experimental observation of Bisogni $et$ $al$~\cite{Bisogni16}.
Thus RIXS can be used to study reconstruction of low-energy electron-hole continuum due to metal-insulator transitions and LDA+DMFT provides an accurate description of it.

In conclusion, we have presented a computational study of XAS and RIXS across the CD/metal-insulator transition in LuNiO$_3$. 
Our results show that the two peaks present
in XAS spectra and reflected in RIXS of the CD insulating phase
cannot be uniquely associated with the LB and SB sites. While the low-energy peak
originates from the LB site, the high-energy peak combines signals from both sites.
The comparison of RIXS spectra in the metallic and insulating phases shows that
subtle changes of the FL feature at low $\omega_{\rm{loss}}$ can be used to identify the metal-insulator transition as proposed in Ref.~\cite{Bisogni16}. 


\begin{acknowledgments}
The authors thank V. Bisogni, T. Uozumi and K. Yamagami for valuable discussions. 
A.H, M.W, and J.K are supported by the European Research Council (ERC)
under the European Union's Horizon 2020 research and innovation programme (Grant Agreement No.~646807-EXMAG).
The 
calculations were performed at the Vienna Scientific Cluster (VSC).
\end{acknowledgments}

\bibliography{lunio3_rixs}

\end{document}